# Simultaneously realizing thermal and electromagnetic cloaking by multi-physical null medium

Yichao Liu, Xiaomin Ma, Kun Chao, Fei Sun*, Zihao Chen, Jinyuan Shan, Hanchuan Chen, Gang Zhao, and Shaojie Chen

*Key Lab of Advanced Transducers and Intelligent Control System, Ministry of Education and Shanxi Province, College of Physics and Optoelectronics, Taiyuan University of Technology, Taiyuan, 030024 China*

* Corresponding author: sunfei@tyut.edu.cn

**Simultaneously manipulating multiple physical fields plays an important role in the increasingly complex integrated systems, aerospace equipment, biochemical productions, etc. For on-chip systems with high integration level (e.g., electronic/photonic chips and radio-frequency/microwave circuits), where both electromagnetic information/energy transporting and heat dissipation/recovery need to be considered, the precise and efficient control of the propagation of electromagnetic waves and heat fluxes simultaneously is particularly important. In this study, we propose a graphical designing method based on thermal-electromagnetic null medium to simultaneously control the propagation of electromagnetic waves and thermal fields according to the pre-designed paths. A thermal-electromagnetic cloak, which can create a cloaking effect on both electromagnetic waves and thermal fields simultaneously, is designed by thermal-electromagnetic surface transformation and verified by both numerical simulations and experimental measurements. The thermal-electromagnetic surface transformation proposed in this study provides a new methodology for simultaneous controlling on electromagnetic and temperature fields, which can be used to realize a series of novel thermal-electromagnetic devices such as thermal-electromagnetic shifter, splitter, bender, multiplexer and mode converter. The designed thermal-electromagnetic cloak opens up new ways to create a concealed region where any object within it does not create any disturbance to the external electromagnetic waves and temperature fields simultaneously, which may have significant applications in improving thermal-electromagnetic compatibility problem, protecting of thermal-electromagnetic sensitive components, and improving efficiency of energy usage for complex on-chip systems.**

Simultaneous management of thermal fields and EM waves is crucial for electronic systems in many application fields, such as shielding heat flow on temperature sensitive components in electronic chips[1], cooling in active phased array radar system[2] or high-power laser slab[3], recovery of waste heat in electromagnetic devices[4,5], and simultaneous improvement of electromagnetic harvesting and surface heat collection in solar cells[6]. Especially for electronic/photonic on-chip systems[7-12], different modules are highly integrated in the same area to provide more functionality, faster processing power, lower cost and energy consumption, which will inevitably bring some other problems, such as the electromagnetic compatibility of different modules and the heat dissipation of resistive elements or high-speed computing modules. However, there is still a lack of effective method about the multi-physics control of heat flow and electromagnetic waves on chips, partially due to the lack of proper multi-physics metamaterials that can perform as desired function for both thermal and electromagnetic fields, and partly due to the lack of rigorous and systematic theories to guide how to control of thermal-electromagnetic fields simultaneously. Therefore, a rigorous and effective method that can control thermal-electromagnetic fields simultaneously for on-chip systems are highly required, which can simultaneously solve the problems of electromagnetic compatibility and heat dissipation caused by the increased chip integration level.

Transformation optics/thermodynamics[13-15] is a rigorous theory and powerful design method[16-18] for a single physical field (i.e., EM field or thermal field). Some novel on-chip optical components, e.g., multimode waveguide bender and multimode waveguide crossing, have been theoretically studied by transformation optics[19] and experimentally demonstrated[20-22]. However, all these novel on-chip components are confined in a single physical field, which cannot solve the problem of heat dissipation and electromagnetic compatibility on highly integrated chips simultaneously. With the development of metamaterials[23] and transformation optics, simultaneously manipulating two physical fields with a single device becomes feasible, most of which are for the simultaneously control on static electric fields and temperature fields, e.g., thermal-electrostatic cloaking[24,25], thermal-electrostatic concentrator[26], thermal-electrostatic camouflage[27,28], and bi-functional thermal-electrostatic devices[29,30]. In recent years, researches on multi-physical field control are gradually expanding from thermal-electrostatic fields to other physical fields, such as carpet cloak for electromagnetic, acoustic and water waves simultaneously[31], magnetostatic-acoustic cloak[32], and electromagnetic-acoustic stealth coats[33,34]. However, current multi-physics devices for other physical fields are mainly developed by designing and optimizing metamaterials/metasurfaces for a specific function case by case[31-34], therefore it still lacks a general theory (e.g., transformation optics[24,29] or solving Laplace's equation[25,27,30] in the design of thermal-electrostatic devices) to achieve multi-physical field control on other physical fields. Recently, a surface-transformation method is proposed for the design on various acoustic-electromagnetic devices that can be implemented using copper plate array in air[35], which may provide new perspectives on theoretical approaches to design multi-physical devices for other physical fields. However, there is still no general theory that can be utilized to control EM waves and temperature fields simultaneously, nor is there any relevant report that can produce cloaking effect for EM waves and temperature fields simultaneously.

In this study, we propose a thermal-electromagnetic null medium (TENM) theoretically, which performs as a perfect 'endoscope' for EM waves and thermal fields simultaneously. Then, a surface-designing method, i.e., thermal-electromagnetic surface transformation, is proposed based on the ideal projection feature of TENM. Many thermal-electromagnetic devices of various functions (e.g., thermal/EM splitting, bending, and expanding) can be designed by a graphical way. Thereafter, the method of implementing reduced TENM, which retains the same directional projection properties as ideal TENM, is proposed by staggered copper and expanded polystyrene (EPS) boards. As an example, a thermal-electromagnetic on-chip cloak that can work for both EM waves and thermal fields simultaneously is designed based on proposed thermal-electromagnetic surface transformation, which can protect thermal sensitive electrical components on the chip from the surrounding heat flows while not affecting the EM radiation patterns from radiating components. Finally, a thermal-electromagnetic on-chip cloak is fabricated by staggered copper and EPS on a thin foam sheet covered with thermal pads, whose thermal and electromagnetic properties are experimentally measured by IR camera and vector network analyzer (VNA), respectively, which show expected cloaking effect for both electromagnetic waves and heat fluxes.

Figure 1 illustrates the role of the thermal-electromagnetic cloak designed in this study for an on-chip system with high integration level. As shown in Fig. 1(a), a processing unit may be influenced (or even burned) by the waste heats from the surrounding resistive elements on chip, which may also interfere EM radiation from the surrounding radiating components on chip. Removing the resistive elements and radiating components away from the processing unit can solve this problem, however it reduces integration level of the on-chip system. A thermal-electromagnetic on-chip cloak in Fig. 1(b) can solve the above problem without reducing the integration level: the EM signals and waste heats can be simultaneously guided around the processing unit by the designed cloak. As a result, the waste heats are collected by cooling/recovery units and the radiation pattern produced by radiating components is not affected by other neighboring components on the chip, which can effectively solve both electromagnetic compatibility problem and heat dissipation/recovery problem associated with increased integration level for on-chip systems.

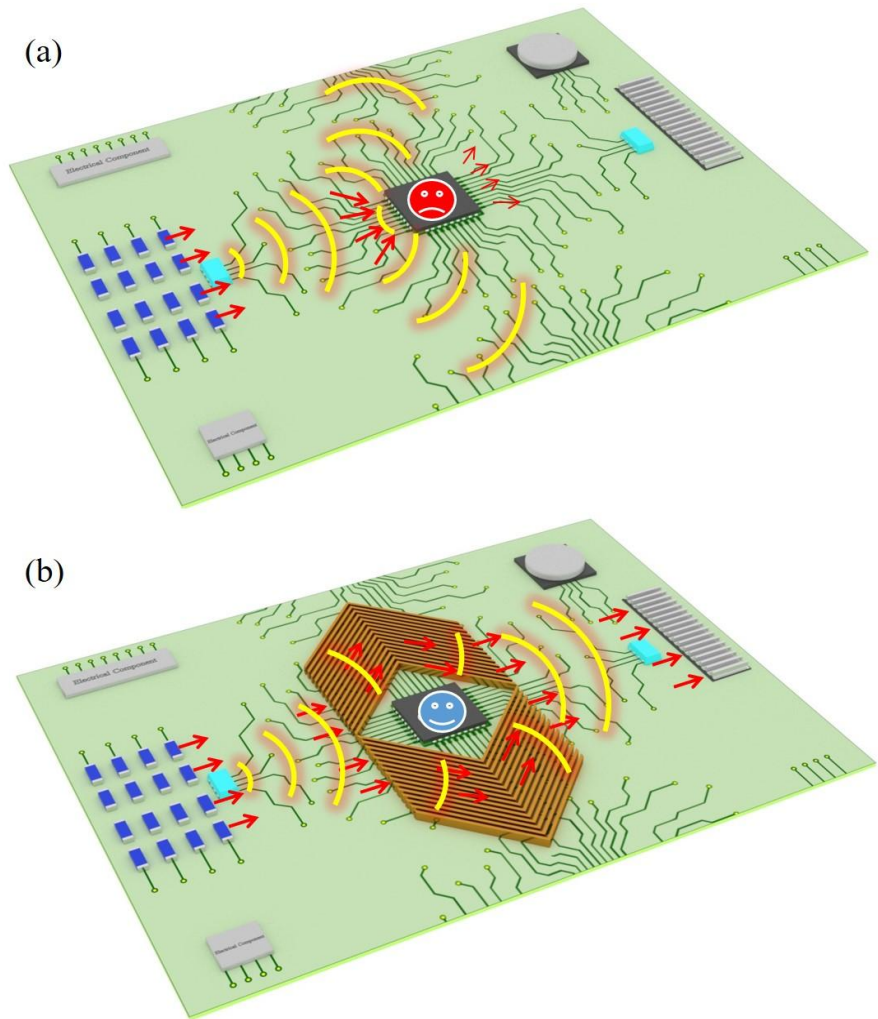

Fig. 1. Schematic diagram of an on-chip system consisting of multiple functional modules (a) without the cloak and (b) with the designed thermal-electromagnetic cloak, respectively. (a) A central processing unit CPU (i.e., thermal sensitive electrical element in the center region) will be affected by the gathered waste heat (indicated by red arrows) from surrounding resistive elements (blue blocks) on chip, which may make the temperature around the processing unit higher than its rated temperature or generate thermal stress/deformation due to the gradient temperature field, and then affect its working efficiency and aggravate the aging. At the same time, the CPU will disturb the EM signals (represented by yellow curves) from surrounding radiating components (e.g., the cyan antenna). (b) The designed thermal-electromagnetic cloak (colored orange) is set around the CPU in the same on-chip system. In this case, the EM signals and waste heats can be simultaneously guided around the thermal sensitive CPU. As a result, the CPU will not be affected by waste heats (or the gradient temperature field from surrounding resistive elements) and not influence radiation pattern of EM signals from surrounding radiating components. Meanwhile, the waste heat can be effectively collected by the latter cooling/recovery units.

**Multi-physical null medium for both electromagnetic waves and thermal fields**

Mono-physical null medium (also referred as to nihility/void medium, or transformation-invariant material), such as optic-null medium for EM waves [36-40], acoustic-null medium for acoustic waves [41-43], magnetic hose for magnetostatic fields[44], and thermal hose for thermal fields [45,46], provides a flexible and simple way to control single physical field due to the nature of its perfectly oriented projection. For example, optic-null medium is a kind of highly anisotropic EM medium whose permittivity and permeability are infinitely large along its principal axes and close to zero in other directions, which can guide EM field along its principal axes and perform as a perfect 'endoscope' for EM field [36-40]. However, there is still a lack of theoretical and experimental studies on multi-physical null medium, e.g., thermal-electromagnetic null medium (TENM) that performs as a perfect 'endoscope' for EM waves and thermal fields simultaneously. In this study, we will show how to derive the required parameters of TENM by an extreme stretching in transformation optics/thermodynamics, and realize the reduced TENM by natural materials.

Using transformation optics and transformation thermodynamics, the three material parameters (i.e., permittivity, permeability, and thermal conductivity) in the physical space (without prime) and the reference space (with prime) can be established through the following relationships [13-15]:

$$\alpha = \frac{J\alpha' J^T}{\det(J)}, \tag{1}$$

where $\alpha$ (and $\alpha'$) can be the relative permittivity $\varepsilon$, the relative permeability $\mu$, or the thermal conductivity $\kappa$, and $J = \partial(x, y, z)/\partial(x', y', z')$ is the Jacobian matrix representing the coordinate transformation between two spaces.

TENM can have flexible shapes (see Fig. 2(a)), which performs as a perfect 'endoscope' linking two arbitrarily shaped surfaces $S_1$ and $S_2$. To derive the material parameters of the TENM, the whole region of arbitrary shape filled with TENM can always be divided into many consecutive small trapezoids (some of them are randomly indicated by small orange trapezoids in Fig. 2(a)) by this way: once the two end faces $S_1$ and $S_2$ of an arbitrarily shaped TENM are determined, a one-to-one correspondence can be established by geometrically projecting the points on $S_1$ and $S_2$ through the blue curve segments in Fig. 2(a), and the adjacent curve segments are separated by a sub-wavelength distance $H$; the region enclosed by adjacent blue curve segments can be further decomposed into several small trapezoids, whose width $D$ is much smaller than the wavelength (see Fig. 2(b)). Next, we only need to calculate the material parameter of each small trapezoid by transformation optics/thermodynamics. The top and bottom sides of each small trapezoid are parallel to the local principal axis of the TENM, which is labeled as the $x$-axis of the local coordinate system in Fig. 2(b).

Firstly, we consider the coordinate transformation that can transform a thin slab in the reference space (see Fig. 2(c)) to a small trapezoid in the physical space (see Fig.

2(b)), which can be written as:

$$\begin{cases} x = M(y)x' + X_0(y) \\ y = y' \\ z = z' \end{cases}, \qquad (2)$$

where the stretching factor $M$ is defined as $M(y) = \dfrac{D + y(\cot\theta_2 - \cot\theta_1)}{\Delta}$,

and $X_0(y) = y\cot\theta_1$. $D$ and $\Delta$ are the length of the bottom side of the trapezoid in the physical space and the small slab in the reference space, respectively. $\theta_1$ ($\theta_2$) is the angle between $x$ axis and the left (right) side of the trapezoid in the physical space as shown in Fig. 2(b). The Jacobian matrix for the transformation in Eq. (2) can be written as:

$$J = \begin{pmatrix} M(y') & x'\dfrac{\partial M(y')}{\partial y'} + \dfrac{\partial X_0(y')}{\partial y'} & 0 \\ 0 & 1 & 0 \\ 0 & 0 & 1 \end{pmatrix}. \qquad (3)$$

Through the coordinate transformation in Eq. (3), together with the transformation optics/thermodynamics in Eq. (1), the material parameters ($\varepsilon$, $\mu$, and $\kappa$) in the physical/reference space can be related by:

$$\alpha = \alpha' \begin{pmatrix} \dfrac{D + y(\cot\theta_2 - \cot\theta_1)}{\Delta} + \dfrac{[x'(\cot\theta_2 - \cot\theta_1) + \Delta\cot\theta_1]^2}{\Delta[D + y(\cot\theta_2 - \cot\theta_1)]} & \dfrac{x'(\cot\theta_2 - \cot\theta_1) + \Delta\cot\theta_1}{D + y(\cot\theta_2 - \cot\theta_1)} & 0 \\ \dfrac{x'(\cot\theta_2 - \cot\theta_1) + \Delta\cot\theta_1}{D + y(\cot\theta_2 - \cot\theta_1)} & \dfrac{\Delta}{D + y(\cot\theta_2 - \cot\theta_1)} & 0 \\ 0 & 0 & \dfrac{\Delta}{D + y(\cot\theta_2 - \cot\theta_1)} \end{pmatrix}. \qquad (4)$$

Secondly, we consider the case when each small slab in the reference space is extremely thin (i.e., $\Delta \to 0$). In this case, each small trapezoid in the physical space corresponds to a surface with null volume in the reference space, and hence the corresponding material parameters in each small trapezoid reduces to the TENM (i.e., null medium). Since two sides of each small trapezoid filled with TENM in the real space correspond to the same surface in the reference space, the thermal/EM fields on two sides of each small trapezoid should be the same, which means the function of each small trapezoid filled with TENM is ideally projecting thermal/EM fields from its front side to its back side. Taking the limitation $\Delta \to 0$ (note $0 \leq x' \leq \Delta$) in Eq. (4), the required material parameters of ideal TENM can be obtained:

$$\varepsilon = \mu = \kappa \sim \begin{pmatrix} \infty & 0 & 0 \\ 0 & 0 & 0 \\ 0 & 0 & 0 \end{pmatrix}. \tag{5}$$

As shown in Eq. (5), the ideal TENM in each small trapezoid in the physical space is extremely high anisotropic medium, whose permittivity/permeability/thermal-conductivity are all infinitely large along its local principal axis (i.e., the tangential direction of blue curve segment in Fig. 2(a)) and nearly zero in other directions. For a region of arbitrary shape filled with TENM in Fig. 2(a), it can always be decomposed into several local consecutive small trapezoids, whose function is ideally projecting thermal/EM fields from its front side to its back side, and the material parameters in each small trapezoid can be expressed by Eq. (5) in its local principal coordinate system. The small trapezoids are continuously distributed inside the TENM, i.e., the end of the previous small trapezoid continuously connects to the front of the latter small trapezoid, which means the principal axes of each local TENM are also continuously connected. Therefore, the whole thermal/EM fields on an arbitrarily shaped surfaces $S_1$ will be ideally projected onto another arbitrarily shaped surfaces $S_2$, once the region between these two surfaces are filled with TENM whose principal axes link continuously from $S_1$ to $S_2$.

The ideal TENM can theoretically perform as a perfect 'endoscope' for EM waves and thermal fields simultaneously, whose performance is also verified by numerical simulations in Figs. 2(d) and 2(e). As shown in Figs. 2(d) and 2(e), when an EM/hot line source are placed on one surface $S_1$ of a tubular TENM, both EM wave and heat flux can be directionally guided along the principal axes of the TENM from $S_1$ to $S_2$, which in turn will produce an EM/hot image spot on the other surface $S_2$. The TENM can have a variety of shapes for different applications, such as bifurcated structures for simultaneous splitting of electromagnetic waves and heat flows (see a fractal tree structure in Figs. 2(f) and 2(g)), or porous structures for simultaneous shielding of electromagnetic waves and heat flows (see a 'Tai Chi' shaped structure with a hole in Fig. 2(h) and 2(i)). Note that the required parameters of ideal TENM need infinite and zero values, which can hardly be realized in practice. Later, we will show how to design the reduced TENM that still retains the same directional projection performance as the ideal TENM and can be implemented by natural materials.

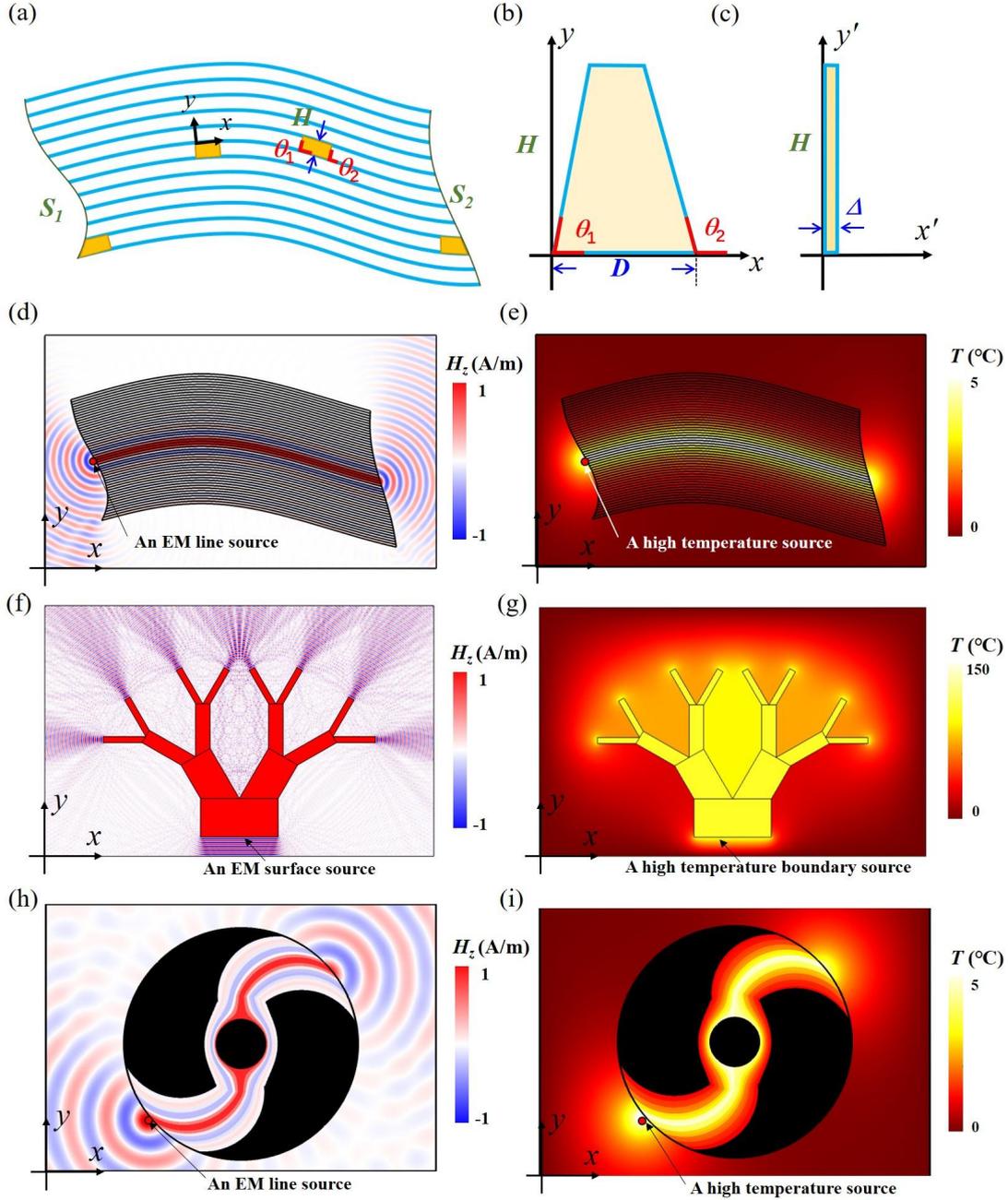

Fig. 2. (a) basic schematic diagram of a tubular TENM that connects two arbitrarily shaped surfaces $S_1$ and $S_2$, which can simultaneously project the thermal-electromagnetic field distribution from $S_1$ onto $S_2$ along its principal axes (blue curves). An arbitrary trapezoid element of a null medium in the (b) physical space is transformed to a compressed thin slab in the (c) reference space. The simulated results when a line current EM source and high temperature source are on the input surface $S_1$ of the TENM, where the normalized magnetic field (d) and the temperature field (e) are plotted, respectively. The simulated normalized magnetic field (f) and temperature field (g) for a fractal-tree shaped TENM with principal axes along the trunk, respectively, which can split both EM fields (f) and heat flux (g) from the root to the top branches. The simulated normalized magnetic field (h) and temperature field (i) for a 'Tai Chi' shaped TENM, respectively, which can guide both EM fields (h) and heat flux (i) around the center concealed hole. The black regions in (h) and (i) represent areas with perfect electric conductor and thermal insulation boundaries that EM fields and heat flux never touch. Details of the numerical setting are given in Supplementary Note 1.

**Thermal-electromagnetic surface transformation by TENM**
Since TENM can project EM and thermal fields simultaneously along its principal axis, we can further propose a surface-designing method, namely thermal-electromagnetic surface transformation, which provide a graphical way to achieve simultaneous control of electromagnetic and thermal fields. With the help of thermal-electromagnetic surface transformation based on TENM, the design problems of thermal-electromagnetic devices can be regarded as a black box problem shown in Fig. 3(a). Once the distribution of the thermal/electromagnetic fields incident to and exiting from the black box are given (e.g., the incoming/outgoing EM wavefront or isothermal surface indicated by the black and red dashed curves in Fig. 3(a)), the process of designing thermal-electromagnetic devices is converted into the problem of graphically determining the shape and principal axis of the TENM inside the black box. The whole graphic design process can be summarized as two steps (see Movie 1): the first step is to determine the shape of the input/output boundaries of the TENM (i.e., black and red solid curves in Fig. 3(a)) according to the incoming/outgoing thermal-electromagnetic fields, which can be designed to be the same as the shape of incoming/outgoing EM wavefront or isothermal surface (e.g., black/red dashed curves have the same shapes as black/red solid curves in Fig. 3(a)). The second step is to determine the principal axes of the TENM inside the black box, which corresponds to the problem of geometrically finding a proper one-to-one projection that can project the points on the TENM's input boundary to the points on the TENM's output boundary (e.g., the blue arrow lines in Fig. 3(a)). In this case, the directions of the geometric projection are exactly the principal axes of the TENM. Note the geometric projection between TENM's input/output boundaries is not unique, which can usually be chosen as a simple linear projection.

A thermal-electromagnetic cloak is designed as an example to illustrate this graphical method as shown in Fig. 3(b). Considering one function of thermal-electromagnetic cloak is to keep the input and output EM wavefront or isothermal surface the same, thus if the input EM wavefront or isothermal surface is planar, the output from the black box should also be a plane. Once the input and output EM wavefront (and isothermal surface) are designed as planes (indicated by the black and red dashed lines, respectively, in Fig. 3(b)), the input boundary (black solid line) and output boundary (red solid line) of the black box can be correspondingly determined as two planes. Considering the other function of thermal-electromagnetic cloak is to create a concealed region where both electromagnetic waves and temperature fields are not subject to any "scattering" by the objects placed inside, as an example, the simplest shape of the concealed region can be designed as a square (indicated by the green region in Fig. 3(b)). The next step to design a thermal-electromagnetic cloak by thermal-electromagnetic surface transformation is "how to project the points from the black solid line segment geometrically one-by-one onto the red solid line segment without touching the green region".

The most convenient geometrical projection is using segmented polylines to connect the input and output surfaces of the cloak, which are indicated by the blue arrowed lines in Fig. 3(b). Note that the directions of arrowed segmented polylines are the same as the principal axes of the TENM filled within the cloak. Once the input and output surfaces of the cloak and the principal axes of the TENM inside the cloak have been determined inside the black box geometrically, the thermal-electromagnetic cloak, which can achieve simultaneous thermal-electromagnetic cloaking of objects in the green region for plane wave incidence, can be obtained by simply filling the region (indicated in yellow) with the TENM whose principal axes are the same as the directions of the blue arrowed segmented polylines. The concealed green region can also be designed as any other shape, just make sure that the projected line segments from the points on the input surface do not touch the concealed region of any other shapes in the projection onto the output surface (i.e., the blue arrowed segmented polylines do not touch the green region; and see examples in Fig. S5 of the Supplementary Note 2).

Figs. 3(e) and 3(f) show the simulated cloaking effect of the thermal-electromagnetic cloak (designed by thermal-electromagnetic surface transformation in Fig. 3(b)) for electromagnetic wave and thermal field, respectively, when the detecting wavefront/isotherm are planes. Considering the perfect projecting feature of the TENM, whatever the form of the detecting wavefronts/isotherms incident on the input surface of the designed cloak, they will be perfectly projected along the principal axis of the TENM onto the output surface of the cloak. Therefore, the designed cloak can still work for any other kinds of detecting wavefronts/isotherms, e.g., a line source in Figs. 3(g) and 3(h). To further compare the cloaking effect, the comparative simulations without the designed thermal-electromagnetic cloak in Figs. 3(e)-(h) are given in Fig. S5 of the Supplementary Note 2.

This graphical method can also be used efficiently in the design for other thermal-electromagnetic devices, such as a thermal-electromagnetic shifter in Fig. 3(c) and a thermal-electromagnetic divider-deflector in Fig. 3(d). The thermal-electromagnetic shifter in Fig. 3(c) can shift the incident thermal-electromagnetic fields simultaneously at a pre-designed distance without changing the direction of the incident thermal-electromagnetic fields. Assuming the incident wavefront/isothermal is a plane (indicated by the black dashed line in Fig. 3(c)) and the outgoing wavefront/isothermal is another plane (indicated by the red dashed line in Fig. 3(c)), which is parallel to the incident plane by a fixed displacement along the $y$ direction. Based on the surface transformation method, the input and output surfaces of the thermal-electromagnetic shifter can be correspondingly determined by using the black and red solid lines in Fig. 3(c), respectively. Then, the principal axes of TENM can be chosen as linear segments that link the input and output surfaces of the designed shifter (indicated by the blue arrowed lines in Fig. 3(c)). Figs. 3(i) and 3(j) show the simulated results for the thermal-electromagnetic shifter in Fig. 3(c), which can shift both thermal-electromagnetic fields by the pre-designed distance along the direction perpendicular to its propagation/diffusion. With the help of the surface transformation method, a structure that can divide and deflect both electromagnetic and thermal fields is shown in Fig. 3(d). Figs. 3(k) and 3(l) show the simulated results for the thermal-electromagnetic divider-deflector in Fig. 3(d), which can divide and deflect both electromagnetic and thermal fields into two parts and propagate/diffuse in different directions.

With the help of the thermal-electromagnetic surface transformation, a series of novel thermal-electromagnetic devices, e.g., thermal-electromagnetic splitter, bender, multiplexer and mode converter, can be designed through the standardized black-box designing steps in Movie 1. More detailed designs, potential applications, and simulation results for other thermal-electromagnetic devices can be found in Supplementary Note 2.

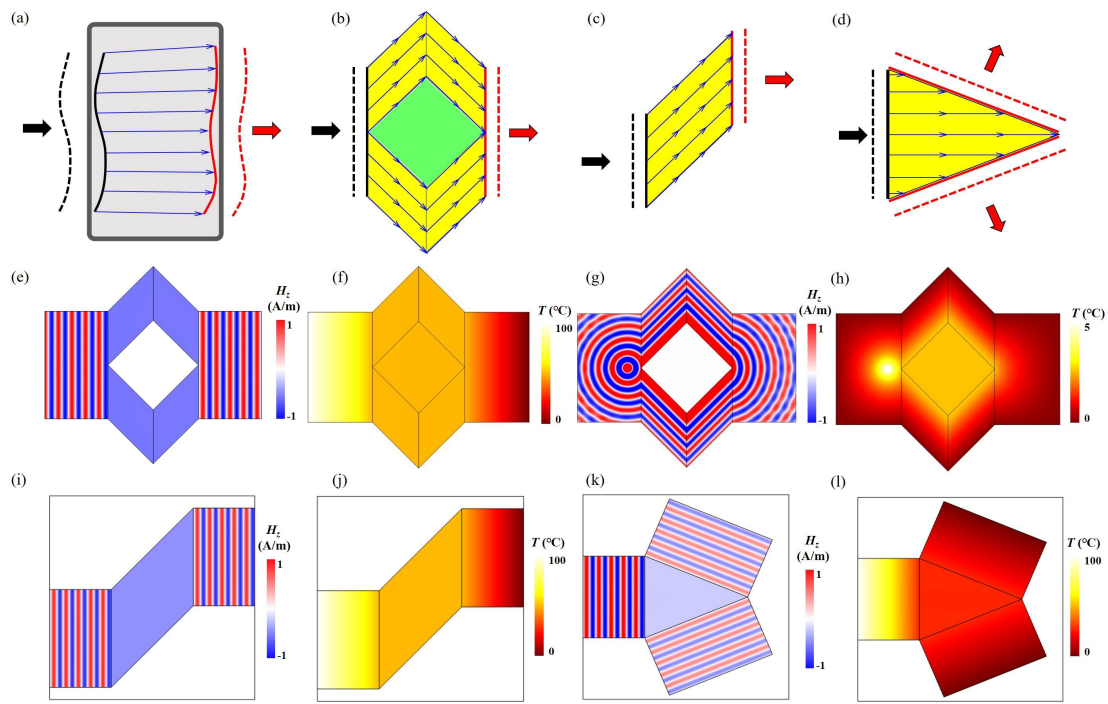

Fig. 3. (a) Schematic diagram of graphical design method based on the directional projection property of TENM. The black and red arrow indicate the input and output fields, and the corresponding wavefronts (or isothermal surface) are represented by black/red dashed curves. Black and red solid curves inside the box are the input/output boundaries of the TENM. Blue arrowed curves indicate one possible projection. (b) A thermal-electromagnetic cloak designed by the graphical method described in (a). The input/output boundaries are designed as straight lines (conformal to the straight input/output EM wavefront and isothermal surface). The yellow region is the TENM with its principal axes along the blue arrowed lines. The green region is the concealed region. (c) A thermal-electromagnetic shifter and (d) a thermal-electromagnetic divider-deflector are designed by the same graphical method described in (a). Simulated magnetic field distributions (e, g) and temperature

distributions (f, h) for the thermal-electromagnetic cloak under the case of a plane detecting wave/isotherm incidence (e, f), and a cylindrical detecting wave/isotherm incidence (g, h). Simulated magnetic field distributions (i, k) and temperature distributions (j, l) for the thermal-electromagnetic shifter (i, j) and divider-deflector (k, l) under a plane wave/isotherm incidence.

**Realizing reduced 2D TENM by staggered copper and EPS boards**

To realize the ideal parameters of TENM in Eq. (5), one effective way is using staggered structure composed by two isotropic media with subwavelength separations. Based on the effective medium theory, the effective electromagnetic/thermal parameters of two staggered isotropic media can be expressed as (see Supplementary Note 3):

$$\begin{cases} \dfrac{1}{\alpha_\perp} = \dfrac{f_1}{\alpha_1} + \dfrac{f_2}{\alpha_2} \\ \alpha_{//} = f_1\alpha_1 + f_2\alpha_2 \end{cases}, \quad (6)$$

where $\alpha$ can be relative permittivity $\varepsilon$, relative permeability $\mu$, or thermal conductivity $\kappa$, and $f_i$ ($i$=1,2) is the filling factor of the $i$-th isotropic medium. $//$ and $\perp$ indicate the directions that are parallel and orthogonal to the interface of two media, respectively. However, to realize the ideal TENM in Eq. (5) by the staggered isotropic media in Eq. (6), it requires one medium with zero parameters (e.g., $\varepsilon_1 = \mu_1 = \kappa_1 = 0$) and the other medium with infinite parameters (e.g., $\varepsilon_2 = \mu_2 = \kappa_2 \to \infty$), which cannot be achieved by natural materials. Since the key to designing various thermal-electromagnetic devices by thermal-electromagnetic surface transformation is to create perfect directional projection on electromagnetic wave and temperature field simultaneously based on the property of TENM, media with simplified material parameters (referred as to reduced TENM), which can also produce the same directional projection on electromagnetic wave and temperature field simultaneously (but with small scatterings), can also be used as basic elements in thermal-electromagnetic surface transformation to design various thermal-electromagnetic devices.

Considering electromagnetic parameters and thermal conductivity of materials in nature from materials handbooks[47-49], we find that staggered copper plates and EPS boards with subwavelength separations in Fig. 4(a) can perform as a reduced 2D TENM that can realize the simultaneous projection of TM-polarized electromagnetic waves and in-plane temperature fields. In this case, the two staggered media in Eq. (6) are copper (with $\varepsilon_1 \to \infty$, $\mu_1$=1, $\kappa_1$=400 W/(m·K), and $f_1$=($p$-$w$)/$p$) and EPS (with $\varepsilon_2$=1, $\mu_2$=1, $\kappa_2$=0.04W/(m·K), and $f_2$=$w/p$), respectively, whose effective electromagnetic/thermal parameters can be calculated by Eq. (6)[50,51]:

$$\begin{cases} \varepsilon_x \to \infty \\ \varepsilon_y = 2 \\ \mu_z = 0.5 \\ \kappa_x \approx 200 \text{W/(m·K)} \\ \kappa_y \approx 0.08 \text{W/(m·K)} \end{cases}, \qquad (7)$$

where the filling factor is chosen as $f_1=f_2=0.5$. As the staggered media performs as a better thermal null medium but a poorer EM null medium for a larger filling factor of copper $f_1$, while the staggered media performs as a better EM null medium but a poorer thermal null medium for a smaller filling factor of copper $f_1$, thus we keep $f_1=f_2=0.5$ in our later design to balance the performance of the reduced TENM for both EM and temperature fields.

To verify the performance of the reduced TENM in Eq. (7), a 2D thermal-electromagnetic shifter designed by thermal-electromagnetic surface transformation in Fig. 3(c), which can shift the distribution of thermal-electromagnetic fields from its input surface to its output surface by a pre-designed distance, can be realized by the staggered copper plates and EPS boards in Fig. 4(b), whose arrangement direction is at a fixed angle 45° with the $x$ axis and the separation of each plate is $w = p/2 = \lambda_0/10$ ($\lambda_0$ is the wavelength of the incident EM wave). Compared with the ideal thermal-electromagnetic shifter in Figs. 3(i) and 3(j), simulated results verify that the 2D thermal-electromagnetic shifter realized by the reduced TENM can still perform satisfied shifting effect for both EM waves (e.g., plane TM-polarized EM wave and line EM source in Fig. 4(c) and 4(e), respectively) and heat flux (e.g., plane isotherm and line heat source in Fig. 4(d) and 4(f), respectively).

The 2D thermal-electromagnetic cloak designed by thermal-electromagnetic surface transformation in Figs. 3(b) can also be realized by the reduced TENM, where the separation between copper plate and EPS board is chosen as $w = p/2 = \lambda_0/11$. The performance of the thermal-electromagnetic cloak by the reduced TENM is verified by numerical simulations in Figs. 4(g)-4(j) for varies thermal-electromagnetic detecting sources. Compared with the ideal thermal-electromagnetic cloak in Figs. 3(e)-(h), the thermal-electromagnetic cloak by the reduced TENM in Figs. 4(g)-4(j) can still provide a satisfied thermal-electromagnetic cloaking effect. There are some small EM reflections/scatterings around the shifter/cloak by the reduced TENM due to the impedance mismatch between air and the reduced TENM, which can be reduced by choosing a lower separation $p$ between two staggered isotropic media or a smaller filling factor of copper $f_1$. More details on the selection of fill factor $f_1$ and bandwidth analysis of the reduced TENM can be found in Supplementary Note 3.

Various other 2D thermal-electromagnetic devices designed by thermal-electromagnetic surface transformation can be realized by the reduced TENM in a similar way. For different applications, 2D thermal-electromagnetic devices can be made/fabricated very high (as long as the incident TM-polarized-wave/isotherm is located within the same 2D cross-section of the device)[40] or made/fabricated within a surface (as long as the incident TM-polarized-wave/isotherm is confined on the same surface) for on-chip thermal-electromagnetic control in Fig. 1. Next, a 2D thermal-electromagnetic cloak by reduced TENM within a surface is fabricated, whose on-chip thermal-EM cloaking performance is verified experimentally.

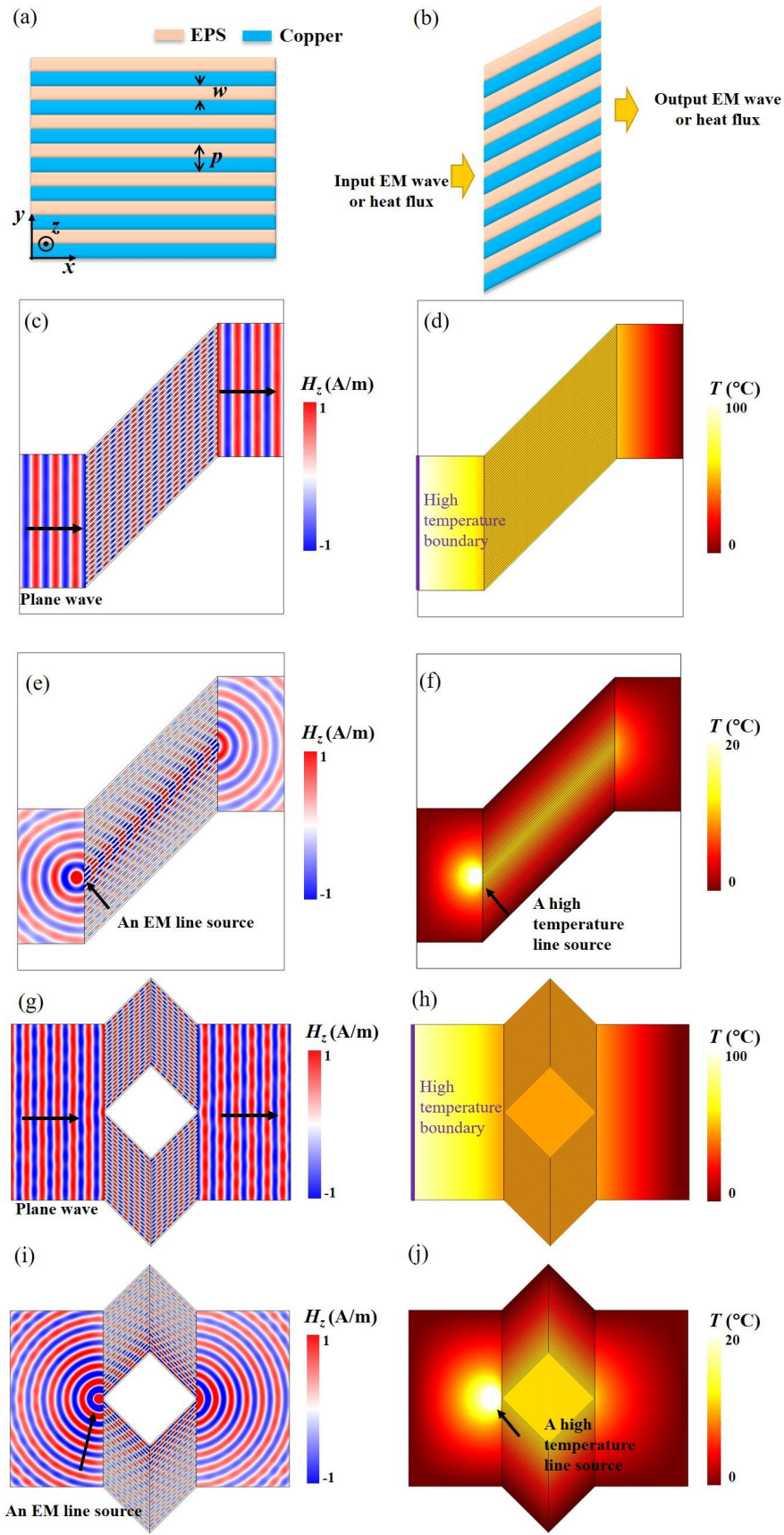

Fig. 4 (a) Reduced TENM by staggered copper and EPS boards. (b) a thermal-electromagnetic shifter realized by reduced TENM. Simulated 2D

magnetic/temperature field distributions for a thermal-electromagnetic shifter with reduced TENM under (c, d) a plane TM-polarized-wave/isotherm incidence and (e, f) a line thermal/electromagnetic source. Simulated 2D magnetic/temperature field distributions for a thermal-electromagnetic cloak with reduced TENM under (g, h) a plane TM-polarized-wave/isotherm incidence and (i, j) a line thermal/electromagnetic source.

**Experimental design and measurements**
To experimentally verify the on-chip cloaking effect of the above designed thermal-electromagnetic cloak by staggered copper plates and EPS boards. A sample cloak in Fig. 5(a) is fabricated by 96 copper plates (colored in orange) and 92 EPS boards (colored in green) with the same size (i.e., length $l = \sqrt{2}a = 141$ mm, width $w$ = 3mm, and height $h$ = 10mm). The working wavelength is designed as $\lambda_0 = 100w/9$ = 100/3mm. The concealed region (i.e., a black protruding square) has a square cross-section in $x$-$y$ plane with diagonal length $2a$ = 200mm and height $h$ = 10mm, which is surrounded by copper plates and filled by EPS. A foam sheet (colored gray with the size 440mm×616mm×100mm and thermal conductivity 0.04W/m/K) covered by thin thermal pads (colored blue with the thickness $\Delta h$ = 2mm and thermal conductivity 13W/m/K) is utilized to mimic an on-chip operating environment. More details about the sample fabrication can be found in the Supplementary Note 4.

The experimental setup to verify the on-chip thermal-electromagnetic cloaking effect of the fabricated sample is shown in Fig. 5(a). For the thermal experiment part, a heating plate with power supply (Silicone Rubber Heater, and colored yellow in Fig. 5(a)) is sticked at the boundary of one thermal pad, which can provide a constant heat power about 5.7W and create the incident heat flux onto the sample. A cooling plate (Peltier Cooler) with power supply is placed at the boundary of the other thermal pad, which performs as a constant low temperature source and maintain 0°C during the experiment. The fabricated cloak is fixed on the center of the on-chip structure (i.e., a foam sheet covered by thermal pads) during the measurement. Another foam board with the same size is covered on the whole structure during the measurement to avoid thermal convection at the top boundary, which can also mimic the package cover of the chip. An IR camera is used to monitor the temperature changes real-timely. After about 4 hours, the temperature distribution captured by the IR camera no longer changes (i.e., whole system reaches a thermally stable state), and then the measured temperature distribution is recorded and shown in Fig.5 (b). The measured result in Fig.5 (b) is consistent well with the 3D simulated result in Fig.5 (d), which shows the fabricated sample can guide heat flux smoothly around the concealed region and keep a low temperature contrast ($\Delta T_{max}$< 0.5 °C in the experiment) inside the concealed region. Both simulated and measured results verify the designed thermal-electromagnetic cloak can protect the concealed region from suffering extreme temperature contrast and avoid affecting the temperature field distribution outside the cloak (i.e., isotherms in thermal pads remain straights).

As a reference, the case when the cloak is removed and replaced by the background thermal pad is also measured in Fig. 5(c) and simulated in Fig. 5(e), which show that the heat flux cannot be smoothly guided away from the black protruding square and a large temperature contrast occurs inside the black protruding square (i.e., $\Delta T_{max} \geq$ 4.6°C in both the 3D simulation and experiment). In this case, if any thermal sensitive element is placed inside the black protruding region, no thermal protection can be obtained (high temperature field gradient in this region will greatly affect the performance of the thermal sensitive element), and the surrounding temperature field is also significantly affected (i.e., isotherms in thermal pads are distorted). More details about the experimental setup and measurement for thermal fields can be found in the Supplementary Note 5.

For the electromagnetic experiment part, the sample and on-chip structure are the same as the thermal experiment. The experimental setup is also shown in Fig. 5(a), which is placed in a small anechoic chamber surrounded by EM pyramid absorber (QYH-J200). A VNA (ROHDE&SCHWARZ ZVL13, working frequency from 5kHz to 13.6GHz), which can perform as both microwave signal source and detector at the designed frequency $f_0$=9GHz, is used to output electromagnetic wave from one port (Port 1) and detect electromagnetic wave by the other port (Port 2) during the electromagnetic measurement. A source loop antenna connected to the Port 1 of VNA by a coaxial cable, which can perform as a 2D magnetic dipole and mimic a radiating component on microwave on-chip system, is fixed at a distance $L = 30$mm away from the front side of the black protruding region on the chip. A detecting loop antenna, which is connected to the Port 2 of VNA by coaxial cable, is movable along the $y$-axis during the measurement and performs as a field detector. Both the source and detecting antenna are central aligned with the cloak along $z$-direction and 5 mm above the surface of the on-chip structure. During the measurement, the detecting loop antenna is kept at the distance $L = 30$mm from the back side of the black protruding region and moves along the $y$-direction at intervals of $0.255\lambda_0$ (=8.5mm), then S21 parameter of the VNA at different locations (marked by the dots in Fig. 5(f)) is measured to obtain the relative magnetic field distribution (the red dots in Fig. 5(h)). The measured magnetic fields in Fig. 5(h) are consistent well with the 3D simulated result in Fig.5 (f), which verifies that the fabricated cloak can guide the TM-polarized electromagnetic waves radiated from an on-chip radiating component (i.e., a source loop antenna here) smoothly around the black concealed region on the chip without disturbing the EM radiation pattern on the other side of the cloak.

As a reference, we also measure and simulate the case when the cloak is removed and only the object in the concealed region (i.e., a square EPS region enclosed by four copper plates) is kept, which are shown in Figs. 5(h) and 5(g) for the measured and 3D simulated cases, respectively. In this case, most of the electromagnetic waves are greatly scattered by the object in the concealed region, which obviously disturbs the original EM radiation pattern on the other side of the cloak. More details about the measurement for electromagnetic fields can be found in the Supplementary Note 6. Both simulated and measured results verify the designed thermal-electromagnetic cloak can guide EM waves and heat fluxes around the concealed region in an on-chip structure (see Movie 2), which may be an effective way to solve electromagnetic compatibility problem and heat dissipation problem simultaneously for highly integrated on-chip systems.

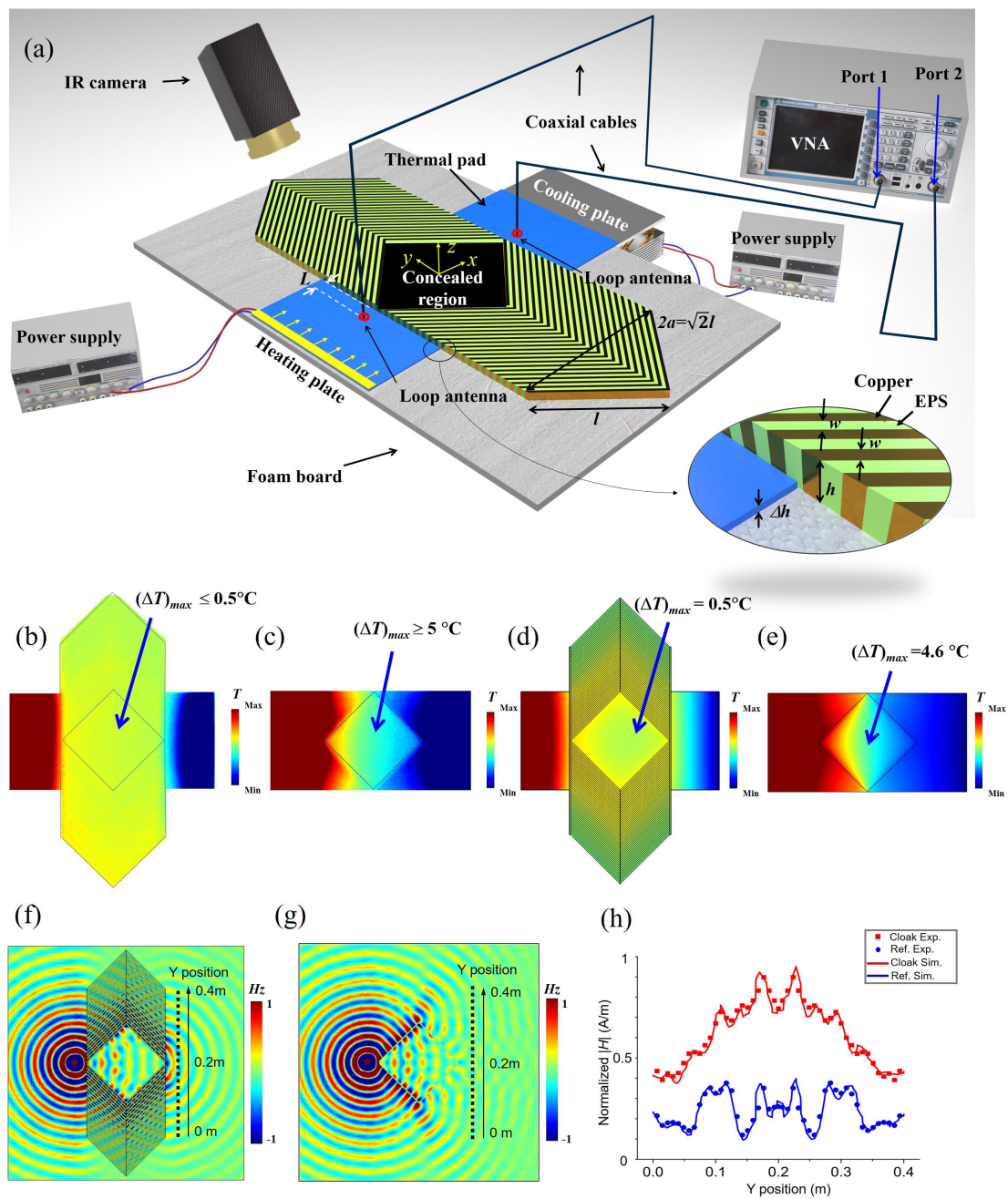

Fig. 5 (a) Schematic of the experimental setup for measuring thermal-electromagnetic cloak. Orange and green slits represent copper and EPS, respectively. Two blue thin thermal pads on the surface of a gray foam board together can simulate an on-chip environment. The black area is the concealed region for both TM-polarized EM waves and heat fluxes. Measured temperature distributions show the black protruding square on the chip will not and will influence the incident plane isothermal with cloak (b) and without cloak (c), respectively. 3D Simulated temperature distributions with cloak

(d) and without cloak (e) are well matched with the measurement results in (b) and (c), respectively. 3D Simulated normalized *z*-component magnetic fields when a radiating component (i.e., a loop antenna) is in front of the black protruding square on the chip with cloak (f) and without cloak (g), respectively. (h) Simulated and measured normalized amplitude of magnetic fields along the marked dotted lines in (f) and (g).

**Conclusion:**
In conclusion, we propose a graphical designing method, i.e., thermal-electromagnetic surface transformation, to simultaneously control the electromagnetic waves and thermal fields. With the help of the proposed method, the design of thermal-electromagnetic device is converted into a design of geometrical projecting problem, and many thermal-electromagnetic devices of various functions can be simply designed through standardized black-box designing steps. In addition, all EM-thermal devices designed by the proposed method, can be realized by staggered copper plates and EPS boards without metamaterials/metasurfaces.

    As an illustrative case, a thermal-electromagnetic on-chip cloak, which can protect thermal sensitive electrical components on the chip from the surrounding heat flows while not affecting the EM radiation patterns from radiating components, is designed and experimentally demonstrated. The simulated and measured results consist very well, which verify that the designed thermal-electromagnetic cloak can guide both EM waves from on-chip antenna and heat fluxes from surrounding thermal components around the concealed region efficiently. The proposed method can also be used to design some thermal-electromagnetic devices of other functions, e.g., simultaneously thermal-electromagnetic field-splitting/multiplexing and mode-converting, which may be used in smart micro controlling system and on-chip system with high integration level.

**Data availability**
The main data and models supporting the findings of this study are available within the paper and Supplementary Information. Further information is available from the corresponding authors upon reasonable request.

**Acknowledgments**
This work is supported by the National Natural Science Foundation of China (Nos. 61971300, 12274317, 61905208), Open Foundation of China-Belarus Belt and Road Joint Laboratory on Electromagnetic Environment Effect (No. ZBKF2022031202), Scientific and Technological Innovation Programs (STIP) of Higher Education Institutions in Shanxi (Nos. 2019L0159 and 2019L0146).

**Author contributions**

Yichao Liu: Methodology, Software, Experiment design, Writing original draft.

Xiaomin Ma: Sample preparing, Thermal and EM measurement, Data recording.

Kun Chao: Thermal measurement, Data recording.

Fei Sun: Conceptualization, Methodology, Supervision, Writing review & editing.

Zihao Chen: Assist in sample preparing, Assist in EM measurement, Data recording.

Jinyuan Shan: Sample preparing, Assist in EM measurement.

Hanchuan Chen, Gang Zhao, Shaojie Chen: Assist in data recording.

All authors contribute to discussions.

**Competing interests**

The authors declare no competing financial interests.